\documentclass[prd,aps,twocolumn,preprintnumbers,showpacs, nofootinbib,superscriptaddress,notitlepage]{revtex4-1}
\usepackage{amsmath,graphicx,epsfig,amssymb}
\usepackage{inputenc}
\usepackage[usenames]{color}
\usepackage{slashed}
\usepackage{multirow}
\usepackage{subfigure}
\usepackage{dsfont}
\usepackage{comment}

\allowdisplaybreaks

\usepackage{floatrow}
\newfloatcommand{capbtabbox}{table}[][\FBwidth]

\begin{document}

\title{$\Xi_c-\Xi_c^{\prime}$ mixing From Lattice QCD }

\author{Hang Liu}
\affiliation{INPAC, Key Laboratory for Particle Astrophysics and Cosmology (MOE),  Shanghai Key Laboratory for Particle Physics and Cosmology, School of Physics and Astronomy, Shanghai Jiao Tong University, Shanghai 200240, China}

\author{Liuming Liu}
\affiliation{Institute of Modern Physics, Chinese Academy of Sciences, Lanzhou, 730000, China}
\affiliation{University of Chinese Academy of Sciences, Beijing 100049, China}

\author{Peng Sun}
\affiliation{Institute of Modern Physics, Chinese Academy of Sciences, Lanzhou, 730000, China}
\affiliation{University of Chinese Academy of Sciences, Beijing 100049, China}

\author{Wei Sun}
\affiliation{Institute of High Energy Physics, Chinese Academy of Sciences, Beijing 100049, China}

\author{Jin-Xin Tan}
\affiliation{INPAC, Key Laboratory for Particle Astrophysics and Cosmology (MOE),  Shanghai Key Laboratory for Particle Physics and Cosmology, School of Physics and Astronomy, Shanghai Jiao Tong University, Shanghai 200240, China}

\author{Wei Wang}
\email{Corresponding author: wei.wang@sjtu.edu.cn}
\affiliation{INPAC, Key Laboratory for Particle Astrophysics and Cosmology (MOE),  Shanghai Key Laboratory for Particle Physics and Cosmology, School of Physics and Astronomy, Shanghai Jiao Tong University, Shanghai 200240, China}
\affiliation{Southern Center for Nuclear-Science Theory (SCNT), Institute of Modern Physics, Chinese Academy of Sciences, Huizhou 516000, Guangdong Province, China}

\author{Yi-Bo Yang}
\affiliation{CAS Key Laboratory of Theoretical Physics, Institute of Theoretical Physics,
Chinese Academy of Sciences, Beijing 100190, China}
\affiliation{School of Fundamental Physics and Mathematical Sciences,
Hangzhou Institute for Advanced Study, UCAS, Hangzhou 310024, China}
\affiliation{International Centre for Theoretical Physics Asia-Pacific, Beijing/Hangzhou, China}
\affiliation{University of Chinese Academy of Sciences, Beijing 100049, China}

\author{Qi-An Zhang}
\email{Corresponding author: zhangqa@buaa.edu.cn}
\affiliation{School of Physics, Beihang University, Beijing 102206, China}

\begin{abstract}
In heavy quark limit, the lowest-lying charmed baryons with two light quarks can form an SU(3) triplet and sextet. The  $\Xi_c$ in the SU(3) triplet and  $\Xi_c'$ in the sextet have the same $J^{PC}$ quantum number and can mix due to the finite charm quark mass and the fact the strange quark is heavier than the up/down quark. We explore the  $\Xi_c$-$\Xi_c'$ mixing by  calculating  the two-point  correlation functions of the $\Xi_c$ and $\Xi_c'$ baryons from lattice QCD. Based on the lattice data, we adopt two independent methods to determine  the mixing angle between $\Xi_c$ and $\Xi_c'$. After making the chiral and continuum extrapolation, it is found that the mixing angle $\theta$ is $1.2^{\circ}\pm0.1^{\circ}$, which seems insufficient to account for the large SU(3) symmetry breaking effects found in weak decays of charmed baryons. 
\end{abstract}

\maketitle

\section{ Introduction} 

Heavy baryons with a  bottom or charm quark provide an ideal place to study the underlying theory for strong interactions. In  heavy quark limit, charmed baryons can be elegantly classified according to heavy quark spin symmetry~\cite{Isgur:1989vq,Isgur:1991wq}. Ground states of  baryons with one heavy quark can be categorized into two sets: if the light quark system is spinless, three baryons form an SU(3) triplet; if the spin of the light quark system is one, then six charmed baryons form a sextet. However in reality, since the charm quark mass is finite and the strange quark is heavier than the up/down quark, charm baryons in the triplet and sextet are likely to mix with each other. An explicit realization is the $\Xi_c-\Xi_c'$ mixing, which is a direct consequence of the effective heavy quark and flavor SU(3) symmetry breaking.

 Weak decays of charmed baryons also contribute to the understanding of the electroweak sector of the standard model. In particular semileptonic decays of charmed baryons are valuable  to extract the CKM matrix element $|V_{cs}|$~(for some recent experimental measurements and lattice QCD calculations , please see Refs.~\cite{BESIII:2015ysy,BESIII:2016ffj,Meinel:2016dqj,Belle:2021crz,Zhang:2021oja} and many references therein). It is interesting to notice that based on  the experimental measurements the flavor SU(3) symmetry, which is a very useful tool widely applied in the analysis of weak decays of heavy mesons,  is found to be sizably broken~\cite{He:2021qnc}. Various mechanisms to explain this observation were explored in Ref.~\cite{He:2021qnc} and a  competitive explanation is through the $\Xi_c-\Xi_c'$ mixing~\cite{Geng:2022yxb}. 
 Very interestingly, several recent works are devoted to determine the mixing angle from various methods, but the results are controversial. Based on experimental data on weak decays of $\Xi_c$ and $\Xi_{cc}$, it is found that a large mixing angle is called for: $\theta=24.66^{\circ}\pm0.90^{\circ}$ in Ref.\cite{Geng:2022yxb,Geng:2022xfz} and $16.27^{\circ}\pm2.30^{\circ}$ in Ref.\cite{Ke:2022gxm}.  A direct calculation in QCD sum rule gives $\theta=5.5^{\circ}\pm1.8^{\circ}$ \cite{Aliev:2010ra}, while in heavy-quark effective theory, it is found $\theta=8.12^{\circ}\pm0.80^{\circ}$ \cite{Matsui:2020wcc}. Besides, an earlier lattice QCD exploration indicates a small mixing angle  \cite{Brown:2014ena}.

This paper presents an update-to-date exploration of the $\Xi_c-\Xi_c'$ mixing  from lattice QCD. Using the recently generated lattice configurations, we calculate the matrix elements $\langle 0|O_{\Xi_c^{(\bar{3},6)}}(t)\bar{O}_{\Xi_c^{(\bar{3},6)}}(0)|0\rangle$, where $O_{\Xi_c^{\bar{3}}}$ and  $O_{\Xi_c^{6}}$ denote the interpolators for the anti-triplet and sextet baryon respectively. With the results for the correlation function matrix,  we use the direct fitting method to fit the energy levels and the mixing angle. As a comparison, the lattice spectroscopy method will also be used. After extracting the mixing angles with different lattice spacings and pion masses, we extrapolate the result to the chiral and continuum limit. We finally find that the mixing angle is at the order of  $1^\circ$, which is likely to indicate the necessity to include other SU(3) symmetry-breaking effects in $\Xi_c$ decays.

The rest of this paper is organized as follows. 
The theoretical framework for the $\Xi_c-\Xi_c'$ mixing  is collected in Sec. II.  We present our lattice simulations  of the two-point correlation functions in Sec.III and a  determination of the mixing angle from a direct fit in Sec.~IV.  Sec. V gives an analysis of the mixing angle through solving the generalized eigenvalue problem.  The conclusion of this work is given in Sec. VI.

\section{$\Xi_c-\Xi_c'$ Mixing in Heavy Quark Effective Theory}

In quark model, a baryon is made of three quarks. In heavy quark limit, heavy baryons with one charm quark can be well classified according to the angular momentum $J$ of light quark system. 
If the spin wave function of the light quark system is antisymmetric, i.e. $J_{qq}=0$, the Fermi statistics and antisymmetric feature in color space require that the  flavor wave function is also antisymmetric in $SU(3)_F$. This corresponds  the decomposition $(3\times3)_{\mathrm{antisymmetric}}=\bar{3}$. If $J_{qq}=1$, the $SU(3)_F$ flavor wave function should be symmetric and hence transforms as $(3\times3)_{\mathrm{symmetric}}=6$.


In heavy quark effective theory (HQET)~\cite{Isgur:1989vq,Isgur:1991wq},  the generic interpolating  operator of a  heavy baryon takes  the form $O=\epsilon^{abc}\left(q_1^{Ta}C\Gamma q_2^b\right)\Gamma'\tilde{Q}^c$, where $q_{1,2}$ denotes a light quark, and $\tilde{Q}$ is the heavy quark field in HQET satisfying $\gamma^0\tilde{Q}=\tilde{Q}$.  In this interpolator we have explicitly specified   the color indices $a,b,c$. The transposition $T$ acts on a Dirac spinor,  and $C=\gamma^0\gamma^2$ is the charge conjugation matrix. Summing over the color indices with the totally antisymmetric tensor $\epsilon^{abc}$ results in a gauge invariant operator. The Dirac matrix $\Gamma$ and $\Gamma'$ are related to the internal spin structures of the heavy baryon. For the ground state, the angular momentum of light quark system is either 0 or 1. If $J=0$ corresponds to the current $q_1^TC\gamma_5q_2$, and $J=1$ corresponds to $q_1^TC\vec{\gamma}q_2$. Therefore, the baryonic current $O_1=\left(q_1^TC\gamma_5q_2\right)\tilde{Q}$ corresponds to the spin $1/2$ baryon, and the  current $\vec{O}_2=\left(q_1^TC\vec{\gamma}q_2\right)\tilde{Q}$ contains both spin-$1/2$ and spin-$3/2$ components. Using $\vec{O}_2$, one can construct the spin-$3/2$ component $\vec{O}_2^{3/2}=\vec{O}_2+\frac{1}{3}\vec{\gamma}\vec{\gamma}\cdot\vec{O}_2$ which satisfying the condition $\vec{\gamma}\cdot\vec{O}_2^{3/2}=0$, and  the spin-$1/2$ component ${O}_2^{1/2}=-\frac{1}{3}\vec{\gamma}\vec{\gamma}\cdot\vec{O}_2=\left(q_1^TC\vec{\gamma}q_2\right)\cdot\vec{\gamma}\gamma_5\tilde{Q}$.

As a result, one can obtain the baryonic currents of $SU(3)_F$ eigenstate as~\cite{Grozin:1992td}
\begin{align}
O^{\bar{3}}=\epsilon^{abc}\left(\ell^{Ta}C\gamma_5 s^b\right)P_+\tilde{c}^c, \\ O^{6}=\epsilon^{abc}\left(\ell^{Ta}C\vec{\gamma} s^b\right)\cdot \vec{\gamma}\gamma_5P_+\tilde{c}^c,
\end{align}
with $\ell=(u,d)$. $P^+=(1+\gamma^0)/2$ is the positive parity projector. 

If the finite charm quark mass and  the differences between the strange quark and up/down quark are taken into account, the two $\Xi_c$ states can mix and their mixing effect could be described by a $2\times2$ mixing matrix:
\begin{align}
    \left(
    \begin{array}{c}
       \left|\Xi_c\right\rangle \\ \left|\Xi_c'\right\rangle 
    \end{array}
    \right) = 
    \left(
    \begin{array}{cc}
       \cos\theta & \sin\theta \\ -\sin\theta & \cos\theta
    \end{array}
    \right)
    \left(
    \begin{array}{c}
       |\Xi_c^{\bar{3}}\rangle \\
       |\Xi_c^{6}\rangle 
    \end{array}
    \right),  \label{eq:mixingpatten}
\end{align}
where $\theta$ denotes the mixing angle. The  mass eigenstates are orthogonal: 
\begin{align}
    H_{\mathrm{QCD}}\left|\Xi_c\right\rangle = m_{\Xi_c}\left|\Xi_c\right\rangle, \quad 
    H_{\mathrm{QCD}}\left|\Xi_c'\right\rangle = m_{\Xi_c'}\left|\Xi_c'\right\rangle,
\end{align}
where $m_{\Xi_c}/m_{\Xi_c'}$ denote the physical baryon masses.

\section{Two-point correlation function From Lattice QCD}

\begin{table*}
\begin{tabular}{c c c c c c c c c}
\hline\hline
   Ensemble &  $\beta$ & $L^3\times T$  & $a$ (fm) & $m_l^{\mathrm{b}}$  & $m_s^{\mathrm{b}}$ & $m_c^{\mathrm{b}}$ & $m_{\pi}$   & $N_{\mathrm{meas}}$    \\\hline
  C11P14L &         &   $48^3\times96$   &             &  $-0.2825$ &  $-0.2310$ & $0.4800$ & $135$ & $203\times48$ \\
  C11P22M & 6.20 &  $32^3\times64$   &  $0.108$ &  $-0.2790$ & $-0.2310$ & $0.4800$ & $222$ & $451\times20$ \\
  C11P29S &         &  $24^3\times72$   &              &  $-0.2770$ & $-0.2315$ & $0.4780$ & $284$ & $432\times26$ \\
	\hline
  C08P30S & 6.41 &  $32^3\times96$   &  $0.080$  &  $-0.2295$ & $-0.2010$ & $0.2326$ & $297$ & $653\times26$ \\\hline
  C06P30S & 6.72 &  $48^3\times144$ &  $0.055$  &  $-0.1850$ &  $-0.1687$ & $0.0770$ & $312$ & $136\times80$  \\\hline\hline
  \end{tabular}
\caption{Parameters of the ensembles used in this work, including the gauge coupling $\beta=10/g^2$, the 4-dimensional volume $L^3\times T$, lattice spacing $a$, bare quark masses $m_{l,s,c}^{\mathrm{b}}$, pion mass $m_{\pi}$ and total measurements $N_{\mathrm{meas}}$. The total measurements  are equal to the number of gauge configurations times the measurements from different time slices on one configuration.} \label{tab:configurations}
\end{table*}

The simulations in this work will be performed on the gauge configurations generated by the Chinese Lattice QCD (CLQCD) collaboration with $N_f=2+1$ flavor stout smeared clover fermions and Symanzik gauge action, at three lattice spacings and three pion masses including the physical one.  The detailed parameters are listed in Table \ref{tab:configurations}. Some previous applications of these configurations can be found in Refs.~\cite{Zhang:2021oja,Wang:2021vqy,Liu:2022gxf,Xing:2022ijm}.

In the numerical simulation, to improve the signal-to-noise ratio of the simulation, we generate the $f$-flavored wall source to point sink propagators 
\begin{align}
S_{{w-p}}^f(\vec{x},t,t_0;\vec{p})=\sum_{\vec{x}_s}e^{-i\vec{p}\cdot(\vec{x}-\vec{x}_s)}S^f(\vec{x},t;\vec{x}_s,t_0)
\end{align}
on the Coulomb gauge fixed configurations. $S^f$ denotes the propagator from wall source at $(\sum_{\vec{x}_s}\vec{x}_s,t_0)$ to the point sink $(\vec{x},t)$. We consider the $2\times2$ correlation function matrix 
\begin{align}
	\mathcal{C}(t, t_0)=\sum_{\vec{x}} \left(
    \begin{array}{cc}
       \left\langle O^{\bar{3}}_p(\vec{x}, t)\bar{O}^{\bar{3}}_w(\vec{0},t_0) \right\rangle 
       		& \left\langle O^{\bar{3}}_p(\vec{x}, t)\bar{O}^{6}_w(\vec{0},t_0) \right\rangle \\
        \left\langle O^{6}_p(\vec{x}, t)\bar{O}^{\bar{3}}_w(\vec{0},t_0) \right\rangle 
        	& \left\langle O^{6}_p(\vec{x}, t)\bar{O}^{6}_w(\vec{0},t_0) \right\rangle 
    \end{array}
    \right) \label{eq:correlationfunctionmatrix}
\end{align}
that contains the baryonic currents. The subscript ``$w$" and ``$p$" denote the wall source and point sink. The related two-point functions can be constructed by the light, strange, and charm quark propagators $S_{{w-p}}^{\{l,s,c\}}$ with momentum $\vec{p}=\vec{0}$
\begin{widetext}
\begin{align}
C_{11}(t,t_0)=&-\sum_{\vec{x}} \epsilon^{abc}\epsilon^{a'b'c'} \Big(S_{{w-p}}^l(\vec{x},t,t_0) \Big)_{\alpha\alpha'}^{aa'}
	\Big( C\gamma_5 S_{{w-p}}^s(\vec{x},t,t_0) \gamma_5C \Big)_{\alpha\alpha'}^{bb'} 
	\Big( P_+ S_{{w-p}}^c(\vec{x},t,t_0) P_+ T  \Big)^{cc'}, \\
C_{12}(t,t_0)=&\sum_{\vec{x}} \epsilon^{abc}\epsilon^{a'b'c'} \sum_{i=1,2,3}\Big(S_{{w-p}}^l(\vec{x},t,t_0) \Big)_{\alpha\alpha'}^{aa'}
	\Big( C\gamma_5 S_{{w-p}}^s(\vec{x},t,t_0) {\gamma}^i C \Big)_{\alpha\alpha'}^{bb'} 
	\Big( P_+ S_{{w-p}}^c(\vec{x},t,t_0) {\gamma}^i\gamma_5 P_+ T  \Big)^{cc'}, \\
C_{21}(t,t_0)=&-\sum_{\vec{x}} \epsilon^{abc}\epsilon^{a'b'c'} \sum_{i=1,2,3}\Big(S_{{w-p}}^l(\vec{x},t,t_0) \Big)_{\alpha\alpha'}^{aa'}
	\Big( C\gamma^i S_{{w-p}}^s(\vec{x},t,t_0) \gamma_5C \Big)_{\alpha\alpha'}^{bb'} 
	\Big( P_+ \gamma^i\gamma_5 S_{{w-p}}^c(\vec{x},t,t_0)  P_+ T  \Big)^{cc'}, \\
C_{22}(t,t_0)=&\sum_{\vec{x}} \epsilon^{abc}\epsilon^{a'b'c'} \sum_{i.j=1,2,3} \Big(S_{{w-p}}^l(\vec{x},t,t_0) \Big)_{\alpha\alpha'}^{aa'}
	\Big( C\gamma^i S_{{w-p}}^s(\vec{x},t,t_0) \gamma^jC \Big)_{\alpha\alpha'}^{bb'} 
	\Big( P_+ \gamma^i\gamma_5 S_{{w-p}}^c(\vec{x},t,t_0) \gamma^j\gamma_5 P_+ T  \Big)^{cc'}
\end{align}
\end{widetext}
where the subscript $\alpha,\alpha'$ denote the Dirac indices and superscript $a,b,c,a',b',c'$ denote the color indices. In the above a trace has been taken and we have adopted a unpolarized projector  $T=(1+\gamma^0)/2$. One can generate the wall source propagators at several time slices $t_0$ and then average the two-point functions constructed by these propagators, so that the statistical precisions of numerical results on different ensembles can reach the same level. The explicit statistics on each ensemble, including the number of gauge configurations and the measurements from different time slices on one configuration, are collected in Table \ref{tab:configurations}.

One can extract the mixing parameters from a joint analysis of both diagonal and non-diagonal terms of $\mathcal{C}$. In the correlation functions, one can insert the mass eigenstates $|n\rangle$ and convert them to local matrix elements. The mismatch of flavor basis baryonic currents and mass eigenstates will be related to the mixing angle under the mixing pattern in Eq.(\ref{eq:mixingpatten}). For example, the correlation $C_{11}$ is given as: 
\begin{widetext}
\begin{align}
C_{11}(t,t_0)=&\sum_{\vec{x}}\langle O^{\bar{3}}_p(\Vec{x},t) \bar{O}^{\bar{3}}_w(\Vec{0},t_0)\rangle \notag\\
=&\sum_{n=\Xi_c, \Xi_c', ...}\frac{e^{-m_n(t-t_0)}}{(La)^3(2m_n)}\langle0|O^{\bar{3}}_p(\Vec{0},0)|n\rangle\langle n|\bar{O}^{\bar{3}}_w(\Vec{0},0)|0\rangle \nonumber\\
=&\frac{1}{\left(La\right)^3} \left[ \frac{e^{-m_{\Xi_c}(t-t_0)}}{2m_{\Xi_c}}  \langle0|O^{\bar{3}}_p(\Vec{0},0)|\Xi_c\rangle\langle \Xi_c|\bar{O}^{\bar{3}}_w(\Vec{0},0)|0\rangle   
+   \frac{e^{-m_{\Xi_c'}(t-t_0)}}{2m_{\Xi_c'}}  \langle0|O^{\bar{3}}_p(\Vec{0},0)|\Xi_c'\rangle\langle \Xi_c'|\bar{O}^{\bar{3}}_w(\Vec{0},0)|0\rangle  + \cdots  \right]. 
\end{align}
\end{widetext}

\begin{figure}
\includegraphics[width=0.95\textwidth]{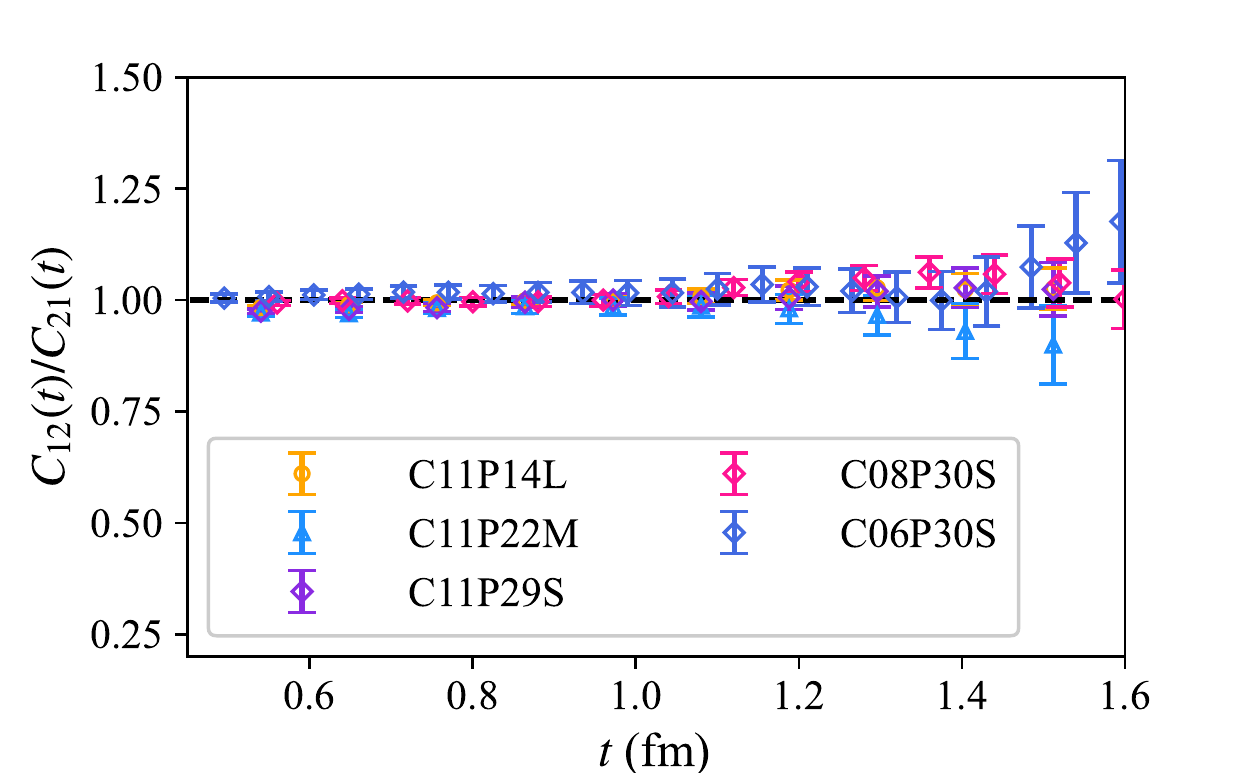}
\caption{Ratios of $C_{12}$ and $C_{21}$ as function of $t$ on different ensembles. Their values are consistent with 1 indicate the lattice artifacts generated by wall source are independent of the mixing in the flavor space.  }\label{fig:ratio1}
\end{figure}

From this expression one can in general determine the mass of the ground state with a large time slice $t$. The excited state contributions are greatly suppressed in this regime, and the correlation functions become
\begin{widetext}
\begin{align}
C_{11}(t,t_0)
=&\frac{1}{\left(La\right)^3} \left[ \frac{e^{-m_{\Xi_c}(t-t_0)}}{2m_{\Xi_c}}  \langle0|O^{\bar{3}}_p(\Vec{0},0)|\Xi_c\rangle\langle \Xi_c|\bar{O}^{\bar{3}}_w(\Vec{0},0)|0\rangle   
+   \frac{e^{-m_{\Xi_c'}(t-t_0)}}{2m_{\Xi_c'}}  \langle0|O^{\bar{3}}_p(\Vec{0},0)|\Xi_c'\rangle\langle \Xi_c'|\bar{O}^{\bar{3}}_w(\Vec{0},0)|0\rangle   \right]  \nonumber\\
=&\frac{1}{\left(La\right)^3} \left[ \frac{e^{-m_{\Xi_c}(t-t_0)}}{2m_{\Xi_c}} \left\langle 0 \left| O^{\bar{3}}_p(\vec{0},0) \right.\right. \left( \left|\Xi_c^{\bar{3}}\right\rangle\cos\theta + \left|\Xi_c^6\right\rangle\sin\theta \right)   \left( \left\langle\Xi_c^{\bar{3}}\right|\cos\theta + \left\langle\Xi_c^6\right|\sin\theta \right) \left| \bar{O}^{\bar{3}}_w(\vec{0},0)\right| 0\right\rangle   \notag\\
&\quad \left. + \frac{e^{-m_{\Xi_c'}(t-t_0)}}{2m_{\Xi_c'}} \left.\left\langle 0 \left| O^{\bar{3}}_p(\vec{0},0) \right.\right. \left(-\left|\Xi_c^{\bar{3}}\right\rangle\sin\theta + \left|\Xi_c^6\right\rangle\cos\theta \right)   \left( -\left\langle\Xi_c^{\bar{3}}\right|\sin\theta + \left\langle\Xi_c^6\right|\cos\theta \right) \left| \bar{O}^{\bar{3}}_w(\vec{0},0)\right| 0\right\rangle \right]  \notag\\
\equiv&A_pA_w^{\dagger} \left[\frac{\cos^2{\theta}}{2m_{\Xi_c}}e^{-m_{\Xi_c} (t-t_0)}+\frac{\sin^2{\theta}}{2m_{\Xi_c'}}e^{-m_{\Xi_c'} (t-t_0)}\right], \label{eq:C11parametrization}
\end{align}
\end{widetext}
where the $t$-independent parameter $A_{p/w}=(La)^{-{3}/{2}}\langle0|O^{\bar{3}}_{p/w}(\Vec{0},0)|\Xi_c^{\bar{3}}\rangle$. In the second step, we have used the mixing matrix in Eq.~\eqref{eq:mixingpatten}. Similarly, the other matrix elements can be expressed as
\begin{widetext}
\begin{align}
C_{12}(t,t_0)=&\sum_{\vec{x}}\langle O^{\bar{3}}_p(\Vec{x},t) \bar{O}^{6}_w(\Vec{0},t_0)\rangle = A_pB_w^{\dagger} \left[\frac{\cos{\theta}\sin{\theta}}{2m_{\Xi_c}}e^{-m_{\Xi_c} (t-t_0)}-\frac{\cos{\theta}\sin{\theta}}{2m_{\Xi_c'}}e^{-m_{\Xi_c'} (t-t_0)}\right], \label{eq:C12parametrization}\\
C_{21}(t,t_0)=&\sum_{\Vec{x}}\langle O^{6}_p(\Vec{x},t) \bar{O}^{\bar{3}}_w(\Vec{0},t_0)\rangle = B_pA_w^{\dagger} \left[\frac{\cos{\theta}\sin{\theta}}{2m_{\Xi_c}}e^{-m_{\Xi_c} (t-t_0)}-\frac{\cos{\theta}\sin{\theta}}{2m_{\Xi_c'}}e^{-m_{\Xi_c'} (t-t_0)}\right], \label{eq:C21parametrization}\\
C_{22}(t,t_0)=&\sum_{\Vec{x}}\langle O^{{6}}_p(\Vec{x},t) \bar{O}^{{6}}_w(\Vec{0},t_0)\rangle = B_pB_w^{\dagger}\left[\frac{\sin^2{\theta}}{2m_{\Xi_c}}e^{-m_{\Xi_c} (t-t_0)}+\frac{\cos^2{\theta}}{2m_{\Xi_c'}}e^{-m_{\Xi_c'} (t-t_0)}\right], \label{eq:C22parametrization}
\end{align}
\end{widetext}
with $B_{p/w}={(La)^{-{3}/{2}}}\langle0|O^{{6}}_{p/w}(\Vec{0},0)|\Xi_c^{6}\rangle$. It should be noticed that the lattice artifacts generated by the wall source and point sink have been collected into the local matrix elements $A_{p/w}$ and $B_{p/w}$, and would not contaminate the energy eigenstates and their mixing. 
Illustrated as Fig.\ref{fig:ratio1}, the ratios $C_{12}/C_{21}$ are consistent with 1 for all the ensembles we used in this calculation, regardless of the pion masses and lattice spacings. Such an observation indicates that the color-spacial factor introduced by the wall source will contribute to an overall factor, which is independent of the mixing in the flavor space.

\section{Mixing Angle from Joint fit}

\begin{table*}
\begin{center}
\renewcommand{\arraystretch}{1.2}
\setlength{\tabcolsep}{1.2mm}
\begin{tabular}{|c|ccccc|}
  \hline
  &$m_{\Xi_c}$ (GeV) & $m_{\Xi_c^{\prime}}$ (GeV) & $\theta$ (${}^{\circ}$) & $\chi^2/$d.o.f  & fit range (fm) \\\hline
  C11P14L& $2.4256(19)$ & $2.5196(22)$ & $1.083(30)$ & $0.96$ &  $1.19-2.81$ \\
  C11P22M& $2.4380(27)$ & $2.5351(30)$ & $0.988(49)$ & $1.0$  & $1.19-2.92$ \\
  C11P29S& $2.4587(27)$ & $2.5536(29)$ & $1.002(50)$ & $1.1$  & $1.19-3.24$ \\
  C08P30S& $2.4753(21)$ & $2.5809(26)$ & $1.080(42)$ & $0.95$ &  $1.20-2.40$ \\
  C06P30S& $2.4695(37)$ & $2.5815(48)$ & $1.021(67)$ & $1.2$ &  $1.32-2.40$ \\ \hline
  Extrapolated &  $ 2.4380 (68)_{\mathrm{stat}}(403)_{\mathrm{syst}}$   &  $2.5562 (74)_{\mathrm{stat}}(422)_{\mathrm{syst}}$   & $1.20 (9)_{\mathrm{stat}}(2)_{\mathrm{syst}}$  &      &      \\ 
  Exp. data \cite{ParticleDataGroup:2022pth}       & $2.46794_{-0.00020}^{+0.00017}$   &   $2.5784\pm0.0005$   &  ---      &  &              \\ \hline
\end{tabular}
\caption{ Results of masses and mixing angles from correlated matrix fits on different ensembles, together with  $\chi^2/$d.o.f and fit ranges $t_{\mathrm{min}}-t_{\mathrm{max}}$ for each fits. The following are extrapolated results, where the experimental data are also listed  for comparison.    }\label{tab:joint-fit}
\end{center}
\end{table*} 

\begin{figure}
\includegraphics[width=0.95\textwidth]{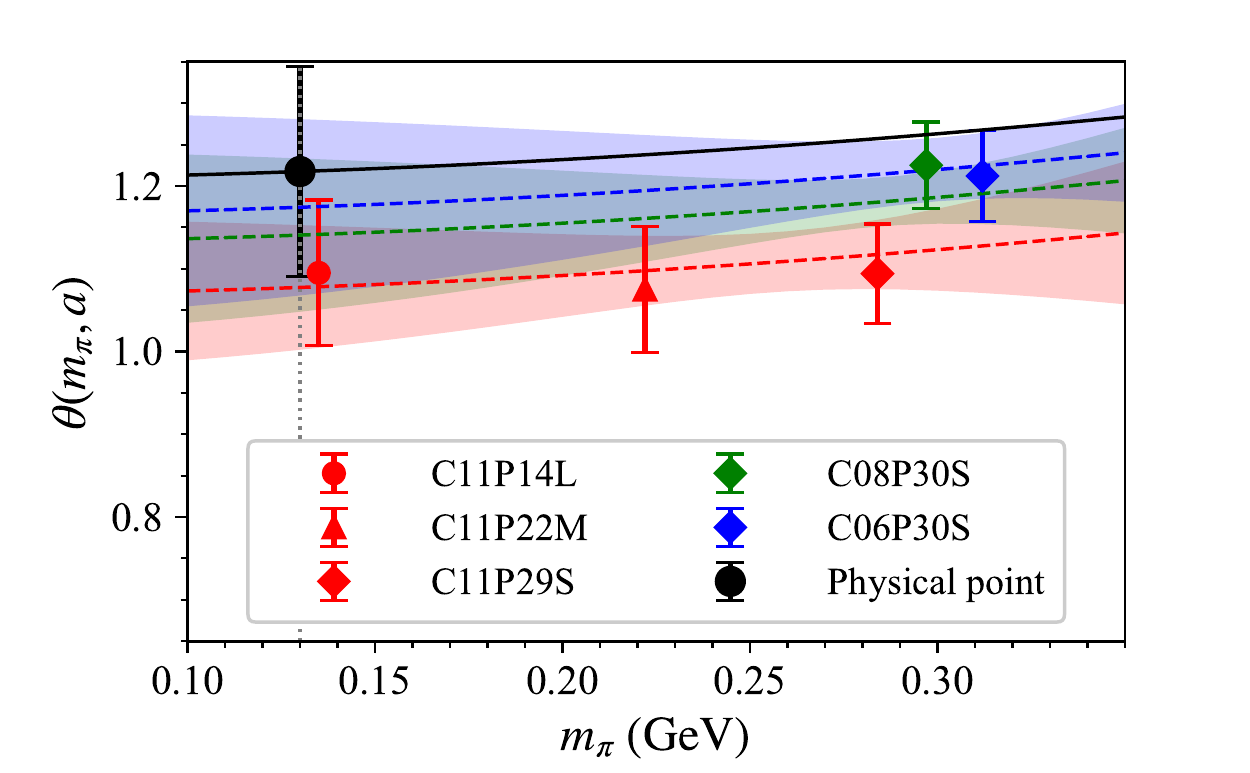}
\caption{Chiral and continuum extrapolations of $\theta$ based on the fit results of Eq.\eqref{eq:extra_formula} as a function of $m_{\pi}$. The extrapolated results from chiral fits at fixed lattice spacings are shown as the red, green, and blue dashed lines with error bands. The black line corresponds to the chiral and continuum extrapolation. The black point denotes the extrapolated values at the physical pion mass and continuum limit. Uncertainties in all data points are statistical only. }\label{fig:theta_from_joint}
\label{fig:spec_all_joint}
\end{figure}

To extract the masses and mixing angle  from the correlation functions, one can apply a joint fit based on the parametrization formula of the correlation function matrix in Eq.(\ref{eq:C11parametrization}-\ref{eq:C22parametrization}) with parameters $A$, $B$, $m_{\Xi_c}$, $m_{\Xi_c'}$ and $\theta$ at $t_0=0$. 

The choice of fit $t$-range relies on both fit quality and physical consideration: the starting time slices $t_{\mathrm{min}}$ after which the data points are included in the fit must be chosen such that contributions from higher excited states have been suppressed sufficiently with the smallest possible statistical uncertainties.  The $t_{\mathrm{max}}$ are determined by avoiding the contaminations from backward-propagating in the periodic boundary condition and avoiding too many degrees of freedom in the $\chi^2$ fit leading to poorly estimating. The optimal choices of the fit range are different for the different ensembles, so we determine them independently to obtain a reasonable description of the joint matrix fit.

\begin{figure}
\includegraphics[width=\textwidth]{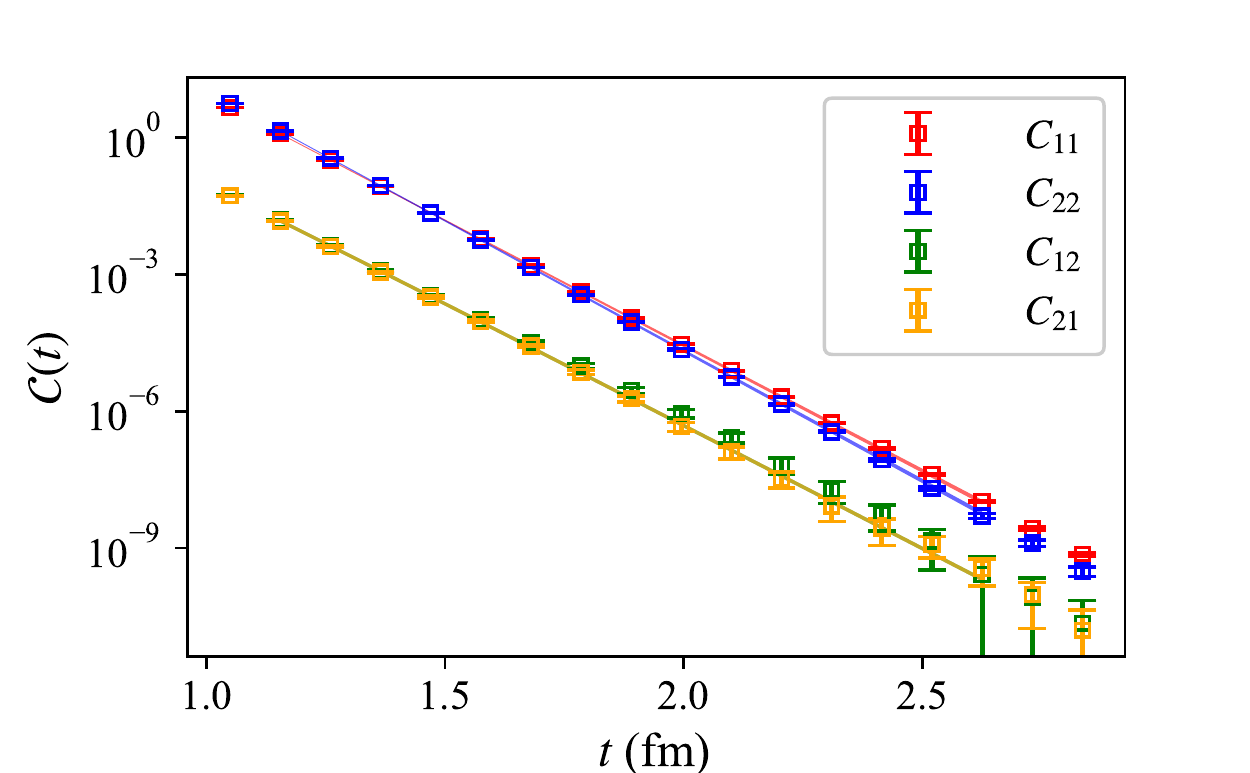}
\caption{Joint fit of the $2\times2$ correlation function matrix using Eq.(\ref{eq:C11parametrization}-\ref{eq:C22parametrization}). The data shown here are from C11P14L, the colored bands indicate the time range and fit results of each matrix element. }\label{fig:correlated_fit}
\end{figure}

Table \ref{tab:joint-fit} includes the fitted values of  $m_{\Xi_c}$, $m_{\Xi_c'}$, $\theta$ and their statistical uncertainties, together with the correlated, reduced $\chi^2$-values ($\chi^2/$d.o.f) and fit ranges $t_{\mathrm{min}}-t_{\mathrm{max}}$ on each ensemble. Of all fits, we evaluate the fit qualities by $\chi^2/\mathrm{d.o.f}\lesssim1$ and the smallest uncertainties for the parameters. As an example, Fig.\ref{fig:correlated_fit} shows the correlated joint fit results on C11P14L, which is perfectly consistent with the data of correlation functions.

After extracting the mixing angles from the ensembles at different pion masses and lattice spacings,  we can  extrapolate these results to the physical values of the quark masses and the continuum limit. We perform the chiral and continuum extrapolation through the following ansatz:
\begin{align}
\theta(m_{\pi}, a)&=\theta_{\mathrm{phy}} + c_1\left( m_{\pi}^2 - m_{\pi, \mathrm{phy}}^2 \right) + c_2 a^2, \nonumber\\
m_n(m_{\pi}, a)&=m_{n,\mathrm{phy}} + c_1\left( m_{\pi}^2 - m_{\pi, \mathrm{phy}}^2 \right) + c_2 a^2 \label{eq:extra_formula}.
\end{align}
These extrapolations are performed at the next-to-leading-order in the chiral expansion. We also included a quadratic dependence of lattice spacings. In order to estimate the systematic uncertainties associated with this truncation, we add the higher-order analytic terms to the fit functions, such as the quartic term $m_{\pi}^4$ in chiral expansion, or $a^3$ term in continuum extrapolation. The latter one, which may arise from heavy-quark discretization errors,  gives sizable   systematic error. In practice, we introduce these terms into the fit functions Eq.(\ref{eq:extra_formula}) separately and calculate the extrapolated masses and mixing angle from the new higher-order fits. The extrapolated results  with both statistic and systematic uncertainties from the higher-order analytic terms are collected in Table \ref{tab:joint-fit}.

With the extrapolation uncertainties taken into account, we find the masses for the $\Xi_c$ and $\Xi_c'$ are both consistent with the experimental data, while the mixing angle is:
\begin{eqnarray}
 \theta= (1.200\pm0.090\pm0.020)^{\circ}. \label{eq:mixing_angle_1}
\end{eqnarray}

\section{Mixing angle from generalized eigenvalue problem}
\label{sec:GEVP}

\begin{figure}
\includegraphics[width=0.95\textwidth]{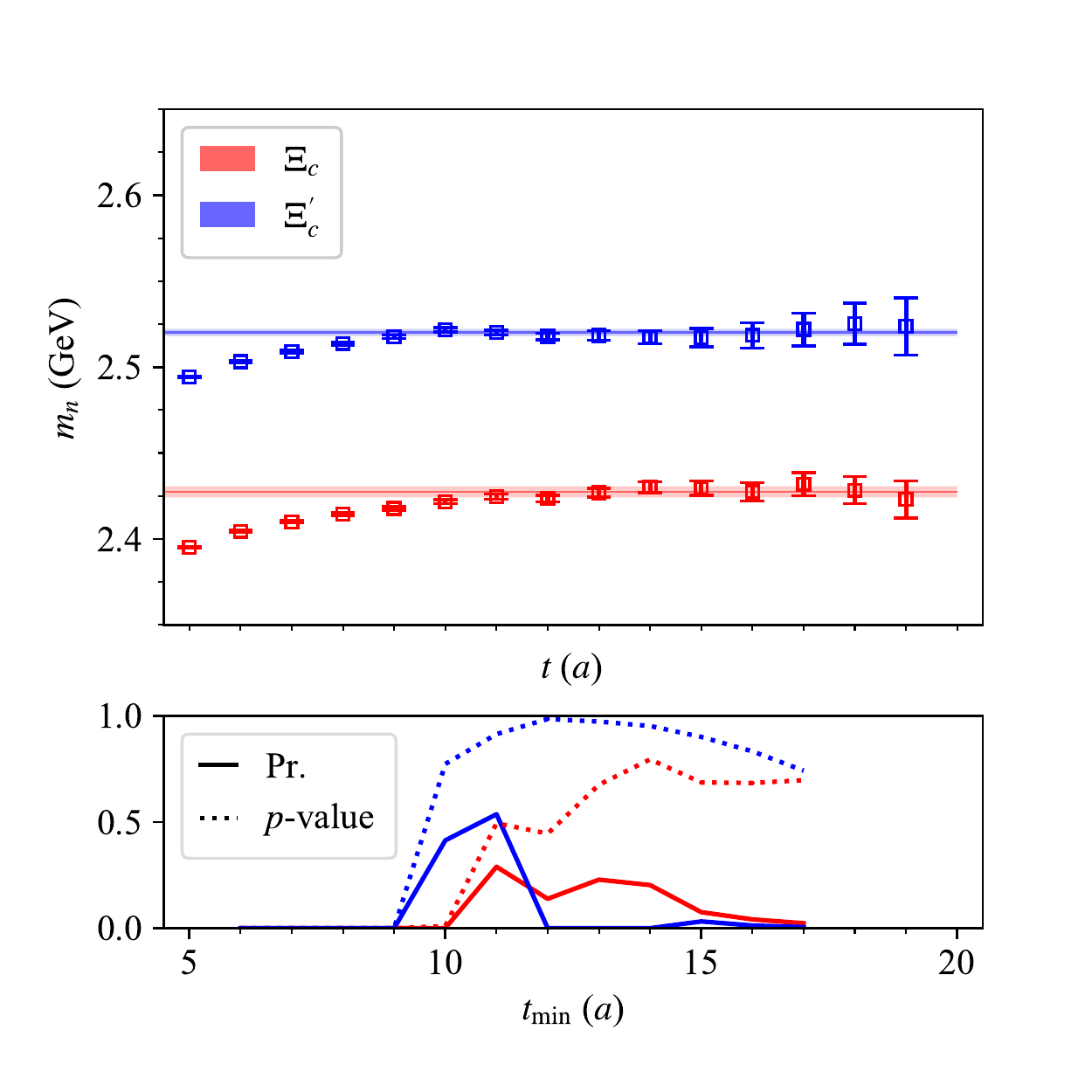}
\caption{Upper panel: effective masses from eigenvalues $\lambda_n(t)$ ($n=\Xi_c,\Xi_c'$),  and model averaging results on C11P14L. The eigenvalues are from solving the GEVP Eq.\ref{eq:gevp} for the correlation function matrix at $t_r=5a$. Lower panel: the model weight factors (solid lines) and standard $p$-values (dashed lines) which reflect the fit quality. }
\label{fig:gevp_effmass}
\end{figure}

Besides the correlated fit, the mixing angle can be also extracted by solving the generalized eigenvalue problem (GEVP) \cite{Michael:1985ne,Luscher:1990ck,Luscher:1990ux,Dudek:2010wm,Blossier:2009kd}
\begin{align}
\mathcal{C}(t)v_n(t)=\lambda_n(t)\mathcal{C}(t_r)v_n(t), \label{eq:gevp}
\end{align}
where $n$ labels the states of $\Xi_c,\Xi_c'$, $\lambda_n(t)$ is the eigenvalue and follows the boundary condition $\lambda_n(t_r)=1$, and there is an orthogonality
condition for the eigenvectors of different states $(n,n')$ as $v_{n'}^{\dagger}\mathcal{C}(t_r)v_n=\delta_{nn'}$. This orthogonality condition allows us to extract the spectrum of near
degenerate states $\Xi_c$ and $\Xi_c'$. The  eigenvalues $\lambda_n$ and eigenvectors $v_n$ usually be solved independently on each time slice $t$ from the source at $t_0$, while for the nearby masses, their eigenvalues might fluctuate time slice by time slice. So we  choose a reference time slice $t_r$ on which reference eigenvector as $v_{n,\mathrm{ref}}$, and compare eigenvectors on other time slices by finding the maximum value of $v_{n',\mathrm{ref}}^{\dagger}\mathcal{C}(t_r)v_n$ which associates a state $n$ with a reference state $n'$.

As discussed above, the two-point correlation function can be decomposed into a time-independent factor and  $e^{-m_nt}$ at large times. In the first factor, the matrix element $\langle0|O^i	|n\rangle$ is related to the eigenvectors which contain the mixing effects between the flavor and energy eigenstates, and the effective masses related to energy eigenstates  can be extracted from the exponential behavior of the correlators $\lambda_n$. The fit function of masses can be expressed as 
\begin{align}
	\lambda_n(t) = c_0 e^{-m_n(t-t_r)} \left(1+c_1e^{-\Delta E(t-t_r)}\right), \label{eq:fitfunctionlambdan}
\end{align}
where the fit parameters  $c_0$ collect the contributions from time-independent factors and $m_n$ denotes the effective mass of $n=(\Xi_c, \Xi_c')$, $c_1$ and $\Delta E$ describe the higher excited states contributions which can be neglected at large time.

\begin{figure}
\includegraphics[width=0.95\textwidth]{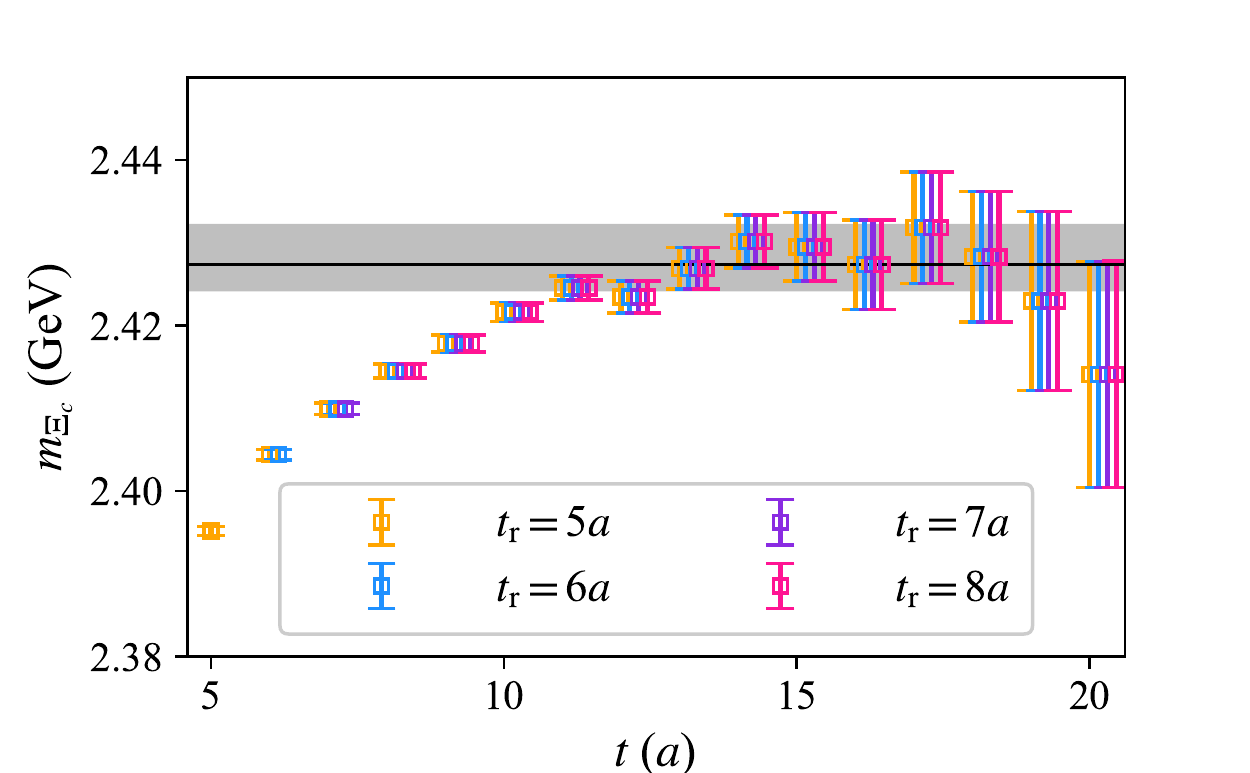}
\caption{ Reference time $t_r$-dependence of $\lambda_{\Xi_c}$ on C11P14L, with $t_r=\{5,6,7,8\}a$. It indicates that the selection of $t_r$ hardly affects the extraction of effective masses. }
\label{fig:gevp_referencetime}
\end{figure}

The extractions of effective masses are illustrated in the upper panel of Fig.\ref{fig:gevp_effmass}. The effective masses $m_n=\log\left( \lambda_n(t)/\lambda_n(t+1) \right)$ are calculated from the eigenvalues of GEVP at reference time slice $t_r=5a$. We also investigate the $t_r$ dependence of GEVP method, shown as Fig.\ref{fig:gevp_referencetime}, we choose several reference time slices to extract and fit the effective masses and find that the selection of $t_r$ hardly affects the results of effective masses. The fit results of masses are collected in Table \ref{tab:gevp}.

In addition, the final results also depend on the fit range. Instead of taking the full difference between the variations of fit ranges as a systematic error as a conservative estimate, it has been advocated that  a technique of Bayesian model averaging~\cite{Jay:2020jkz}  can weight the selections of different fit ranges. This approach allows for a fully rigorous estimation of probability distributions for parameters of interest by combining results from several fits. We adopt this method to investigate the dependence on the fit range in particular the $t_{\mathrm{min}}$. Shown as the lower panel of Fig.\ref{fig:gevp_effmass}, the weight factors $\mathrm{Pr}$ and standard $p$-values describe the fit quality with range start from $t_{\mathrm{min}}$, and  the probability-weighted average of fit results from different $t_{\mathrm{min}}$ will contribute to the final result, which corresponds to the colored band in the upper panel. More details are shwon in the Appendix.

\begin{table*}
\begin{center}
\begin{tabular}{|c|cc|c|}
  \hline
  &$m_{\Xi_c}$ (GeV) & $m_{\Xi_c^{\prime}}$ (GeV)    & $\theta$ (${}^{\circ}$)  \\\hline
  C11P14L& $2.4274(33)$ & $2.5201(20)$  & $1.095 (88)$   \\
  C11P22M& $2.4322(39)$ & $2.5321(44)$  & $1.075 (76)$   \\
  C11P29S& $2.4622(59)$ & $2.5603(79)$  &  $1.094 (60)$  \\
  C08P30S& $2.4763(11)$ & $2.5868(13)$  &  $1.225 (52)$  \\
  C06P30S& $2.4682(38)$ & $2.5905(39)$   &  $1.212 (55)$ \\ \hline
  Extrapolated & $2.428(11)_{\mathrm{fit}}(44)_{\mathrm{ext.}}$ & $2.547(12)_{\mathrm{fit}}(36)_{\mathrm{ext.}}$ &  $1.22(13)_{\mathrm{fit}}(1)_{\mathrm{ext.}}$    \\  
  Exp. data \cite{ParticleDataGroup:2022pth}  & $2.46794_{-0.00020}^{+0.00017}$   &   $2.5784\pm0.0005$  &  ---  \\  \hline
\end{tabular}
\caption{Results of masses and mixing angles from fitting GEVP eigenvalues and eigenvectors with the model averaging approach~\cite{Jay:2020jkz}. The errors from fit contains both static errors and systematic ones associated with choice of fit range, and the second errors denote systematic ones from extrapolation. The extrapolated results and the PDG-averaged ones are also listed here for comparison.   }\label{tab:gevp}
\end{center}
\end{table*} 

\begin{figure}
\includegraphics[width=0.95\textwidth]{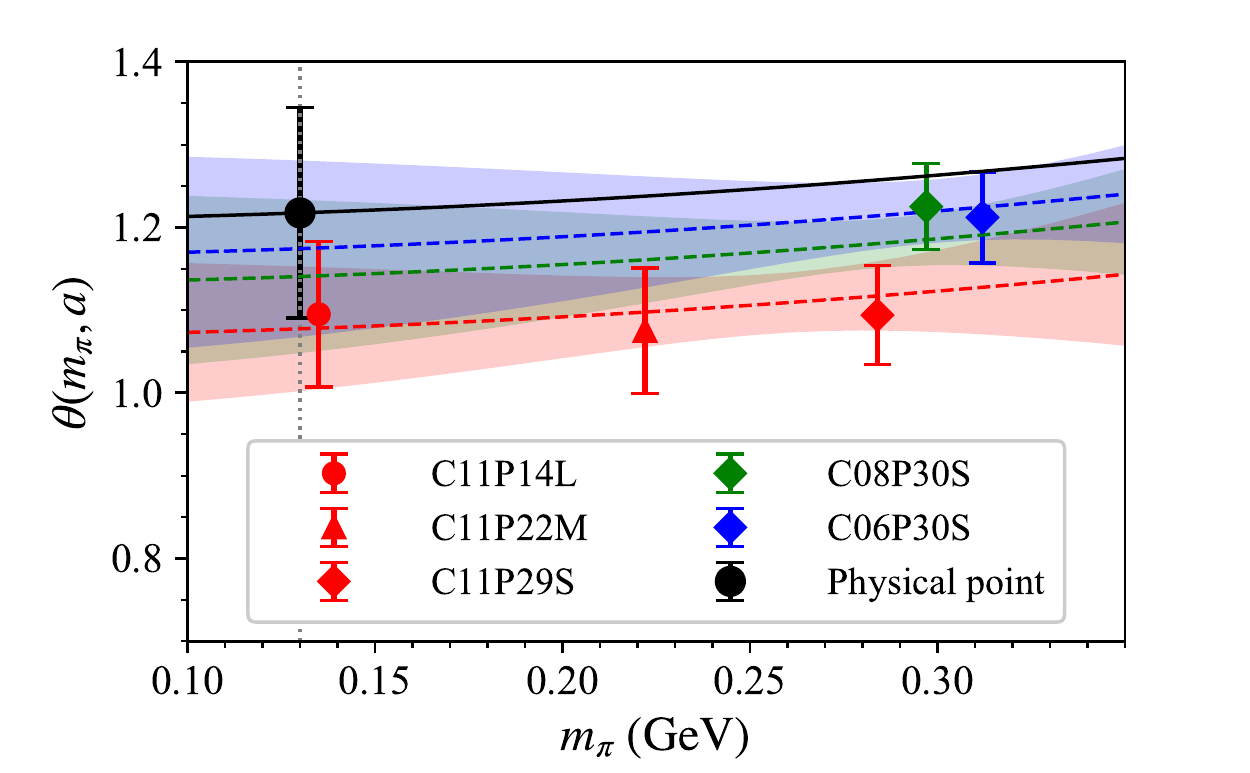}
\caption{Chiral and continuum extrapolations of $\theta$ extracted from GEVP method. The extrapolated results from chiral fits at fixed lattice spacings are shown as the red, green, and blue dashed lines with error bands. The black line corresponds to the chiral and continuum extrapolation. The black point denoted the extrapolated values at the physical pion mass and continuum limit. Errors in all data points are statistical only.   }\label{fig:theta_from_gevp}
\label{fig:theta_extra_gevp}
\end{figure}

After diagonalizing the correlation function matrix $\mathcal{C}$, we find  the GEVP eigenvalues $\lambda_n$ are degenerate to the energy basis, thereby the mixing effects will be collected into the GEVP eigenvectors $v_n$. From the parametrization form in Eq.(\ref{eq:C11parametrization}-\ref{eq:C22parametrization}), the normalized eigenvectors can be expressed as
\begin{align}
	v_1&=\sqrt{1+\frac{A_p^2\cot^2\theta}{B_p^2}}\left(
	\begin{array}{c}
		\frac{A_p}{B_p} \cot\theta \\ 1
	\end{array}
	\right), \\
v_2&=\sqrt{1+\frac{A_p^2\tan^2\theta}{B_p^2}}\left(
	\begin{array}{c}
		-\frac{A_p}{B_p} \tan\theta \\ 1
	\end{array}
	\right),
\end{align}
Therefore the mixing angle $\theta$ can be extracted from $v_{1,2}$.

The fit results on different ensembles are shown in  Tabel \ref{tab:gevp}. As discussed above, we can also extrapolate the mixing angles and masses from different ensembles to their physical values, illustrated as Fig.\ref{fig:theta_extra_gevp}. The comparison of the extrapolated results as well as the PDG averaged ones are also listed in Tabel \ref{tab:gevp}. One can see that the extrapolated results  agree with the ones from the correlated matrix fit.  The estimation of systematic uncertainties is similar to the above discussions.

With the extrapolation uncertainties taken into account, we find the mixing angle is:
\begin{eqnarray}
 \theta= (1.22\pm0.13\pm0.01)^{\circ}, \label{eq:mixing_angle_2}
\end{eqnarray}
which is consistent with Eq.~\eqref{eq:mixing_angle_1} within the uncertainty.  This value is much smaller than the model result extracted  from  experimental data on weak decays~\cite{Geng:2022yxb,Geng:2022xfz,Ke:2022gxm,Xing:2022phq}, and thus insufficient to explain the large SU(3) symmetry breaking effects found in charmed baryon decays. Other mechanisms such as QED corrections should be included.

\section{Charm quark mass dependence}

In HQET, the mixing between $\Xi_c$ and $\Xi_c'$ would vanish in the heavy-quark limit. In this limit, the angular momentum $J$ of the light quark pairs becomes a conserved quantum number, hence the flavor eigenstate which the angular momentum of light degrees of freedom has an unambiguous value would coincide with the energy states. In reality, the mixing occurs through a finite quark mass correction and thus is proportional to $1/m_c$~\cite{Falk:1992wt,Falk:1992ws}.

In this section, we also investigate the charm quark mass dependence of the mixing angle. In practice, we use the ensemble C11P29S to generate the correlation function matrices with several bare charm quark masses around the physical one.  By applying the correlated matrix fits for each case, we obtain the results in Table~\ref{tab:mc_dependence}. 
As one can see, the mixing angle decreases with the increase of charm quark mass, which is within our expectation.

\begin{table*}
\begin{center}
\begin{tabular}{|c|ccccccccc|}
  \hline
  $m_c^{\mathrm{b}}$             &  $0.2$             & $0.3$             & $0.4$              & $0.44$           & $\mathbf{0.478}$          & $0.5$              &        $0.6$             & $0.7$              & $0.8$                  \\\hline
  $m_{\Xi_c}$ (GeV)               &  $2.0987(25)$ & $2.2380(28)$ & $2.3594(26)$ & $2.4069(26)$ & $\mathbf{2.4587(27)}$ & $2.4793(29)$  & $2.5878(30)$ & $2.6898(30)$  &  $2.7859(31)$     \\
  $m_{\Xi_c^{\prime}}$ (GeV) &  $2.1834(24)$ & $2.3249(29)$ & $2.4514(24)$ & $2.4999(24)$ & $\mathbf{2.5536(29)}$ & $2.5718(29)$  & $2.6823(29)$ & $2.7859(30)$   & $2.8835(30)$      \\
  $\theta$ (${}^{\circ}$)            & $1.639(75)$   &  $1.349(73)$  & $1.116(49)$    & $1.049(46)$   & $\mathbf{1.002(50)}$   & $0.969(53)$    & $0.847(47)$   & $0.751(42)$    & $0.674(39)$           \\\hline
\end{tabular}
\caption{ Results of $m_{\Xi_c}$, $m_{\Xi_c'}$ and $\theta$ at different bare charm quark masses. The bold ones correspond to a physical charm quark. These results are extracted from the correlated fit of the correlation function matrix on C11P29S. The total measurements of non-physical mass cases are $N_{\mathrm{meas}}=432\times14$. Uncertainties in all results are statistical only.  }\label{tab:mc_dependence}
\end{center}
\end{table*} 

\begin{figure}
\includegraphics[width=0.95\textwidth]{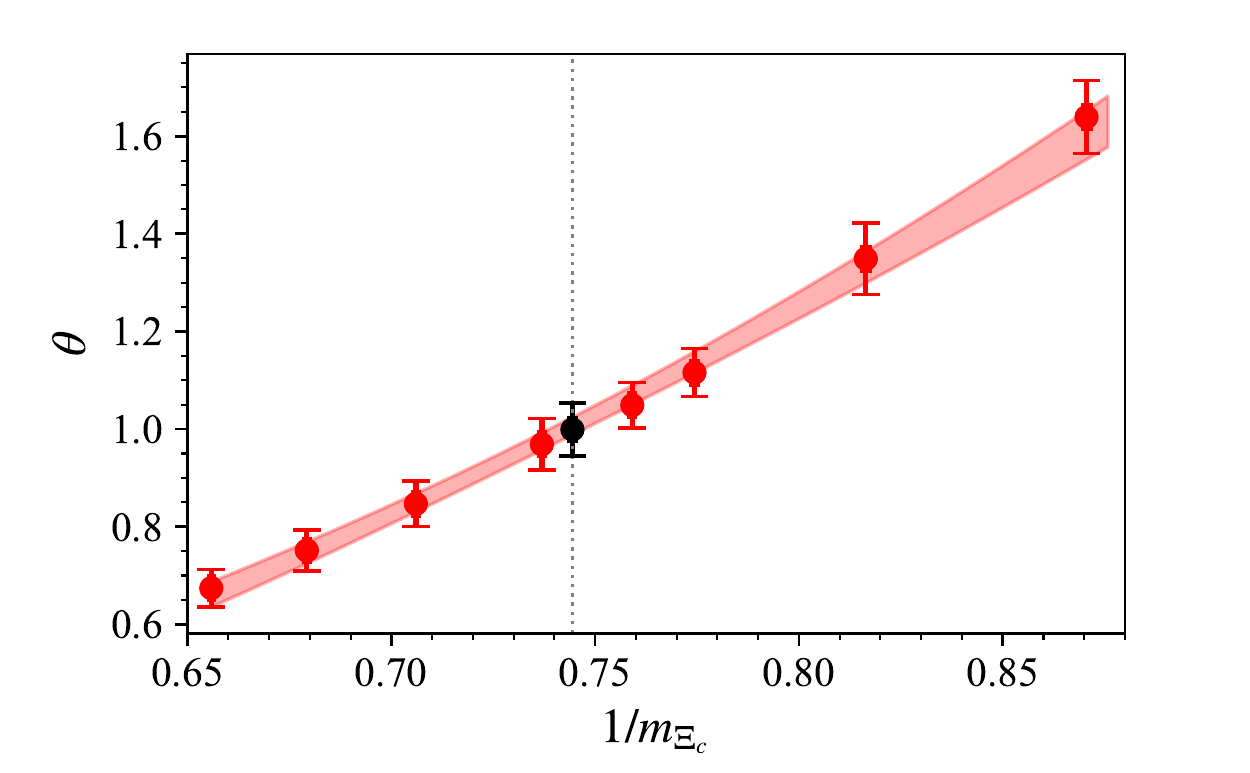}
\caption{Mixing angle $\theta$ as a function of $1/m_{\Xi_c}$. The red data points denote the results based on unphysical charm quark masses, and the black data point denotes the physical one. Uncertainties in all data points are statistical only.  }\label{fig:mc_dependence}
\end{figure}

In order to investigate the heavy quark mass dependence of the mixing angle, and furthermore to predict the behavior at heavy quark limit, we employ a roughly fit ansatz for the mixing angle $\theta$ as a function of $m_{\Xi_c}$
\begin{align}
	\theta = \frac{B_1}{m_{\Xi_c}} + \frac{B_2}{m^2_{\Xi_c}}.
\end{align}
The results obtained from fit are $B_1=-2.78(52)$GeV and $B_2=12.9(1.3)\mathrm{GeV^2}$ with $\chi^2/\mathrm{d.o.f}=0.11$.  It should be noticed that the value of $B_{1,2}$ would suffer from sizeable  discretization effects due to the ensemble we used for this investigation.

\section{Summary}

In summary,  we have explored the  $\Xi_c$-$\Xi_c'$ mixing by  calculating the two-point  correlation functions of the operators to interpolate the $\Xi_c$ and $\Xi_c'$ baryons from the lattice QCD. Based on the lattice data, we have adopted two independent methods to determine  the mixing angle between $\Xi_c$ and $\Xi_c'$.  Direct analysis of the correlation function is conducted and a small mixing angle  is found. This observation is also confirmed through the analysis of solving the generalized eigenvalue problem.

After making the chiral and continuum extrapolation,  we found that the mixing angle $\theta$ is $1.2^{\circ}\pm0.1^{\circ}$, which is significantly small than the ones from other methods. This indicates the $\Xi_c-\Xi_c'$ mixing  is insufficient to account for the large SU(3) symmetry-breaking effects found in weak decays of charmed baryons, and further mechanisms are required. 
A combined investigation of $\Lambda_c$ and $\Xi_c$ decay form factors from lattice QCD that include the mixing effects is undergoing. 

\section*{Acknowledgement}

We are grateful to Fengkun Guo, Xiao-Gang He,  Jiajun Wu, and Qiang Zhao for useful discussions.  We thank the anonymous referee for drawing our attention to the model averaging approach. 
This work is supported in part by Natural Science Foundation of China under grant No. U2032102, 12061131006, 12125503, 12293060, 12293061, 12293062. 
The LQCD calculations were performed using the Chroma software suite \cite{Edwards:2004sx} and QUDA \cite{Clark:2009wm,Babich:2011np,Clark:2016rdz} through HIP programming model \cite{Bi:2020wpt}.
The computations in this paper were run on the Siyuan-1 cluster supported by the Center for High Performance Computing at Shanghai Jiao Tong University, and Advanced Computing East China Sub-center. 

\begin{appendix} 

\section{Model averaging method}

Statistical modeling plays a crucial role in obtaining meaningful results from lattice field theory calculations. While these models are usually rooted in physics, multiple model variations may exist for the same lattice data. To account for the uncertainties associated with model selection,  we employ model averaging \cite{Jay:2020jkz} approach to improve the stability of our results. This approach involves taking a probability-weighted average of all the model variations, providing a more comprehensive and less conservative approach to addressing systematic errors. 

In practice, we consider a set of models $\{M\}$ which fitting all data in the range $[\{\mathrm{list~of~}t_{\mathrm{min}}\}, t_{\mathrm{max}}]$, and combine the statistical results and relevant parameters obtained from fitting in models $\{M\}$ to estimate the the weight factor
\begin{eqnarray}
\mathrm{Pr}(M|D)\approx \rm{exp}\big[-\frac{1}{2}(\chi^2_{aug}(\mathbf{a}^*)+2 k+2N_{cut})\big]
\end{eqnarray}
where $\chi^2_{\rm{aug}}(\mathbf{a}^*)$ represents the standard best-fit augmented chi-squared, $k$ corresponds to the number of fit parameters, and $N_{\rm cut}$ represents the number of removed data points. The normalized $\mathrm{Pr}(M|D)$ are used to estimate the fit qualities of models $\{M\}$. The model-averaged values $\langle a\rangle$ from the fit parameters $\langle a\rangle_M$ can then be determined using the following procedure:
\begin{eqnarray}
   \langle a\rangle=\sum_M \langle a\rangle_M \mathrm{Pr}(M|D), 
\end{eqnarray}
and its error can be estimated by 
\begin{eqnarray}
    \sigma=\sqrt{\sum_M\left[\sigma^2_M+\left(\langle a\rangle_M-\langle a\rangle\right)^2\right]\mathrm{Pr}(M|D)},
\end{eqnarray}
where the $\left(\langle a\rangle_M-\langle a\rangle\right)^2$ term can be interpreted as the variability across the different models.

\begin{figure*}
\centering
\subfigure[ ~	  C11P22M]{\includegraphics[width=0.45\textwidth]{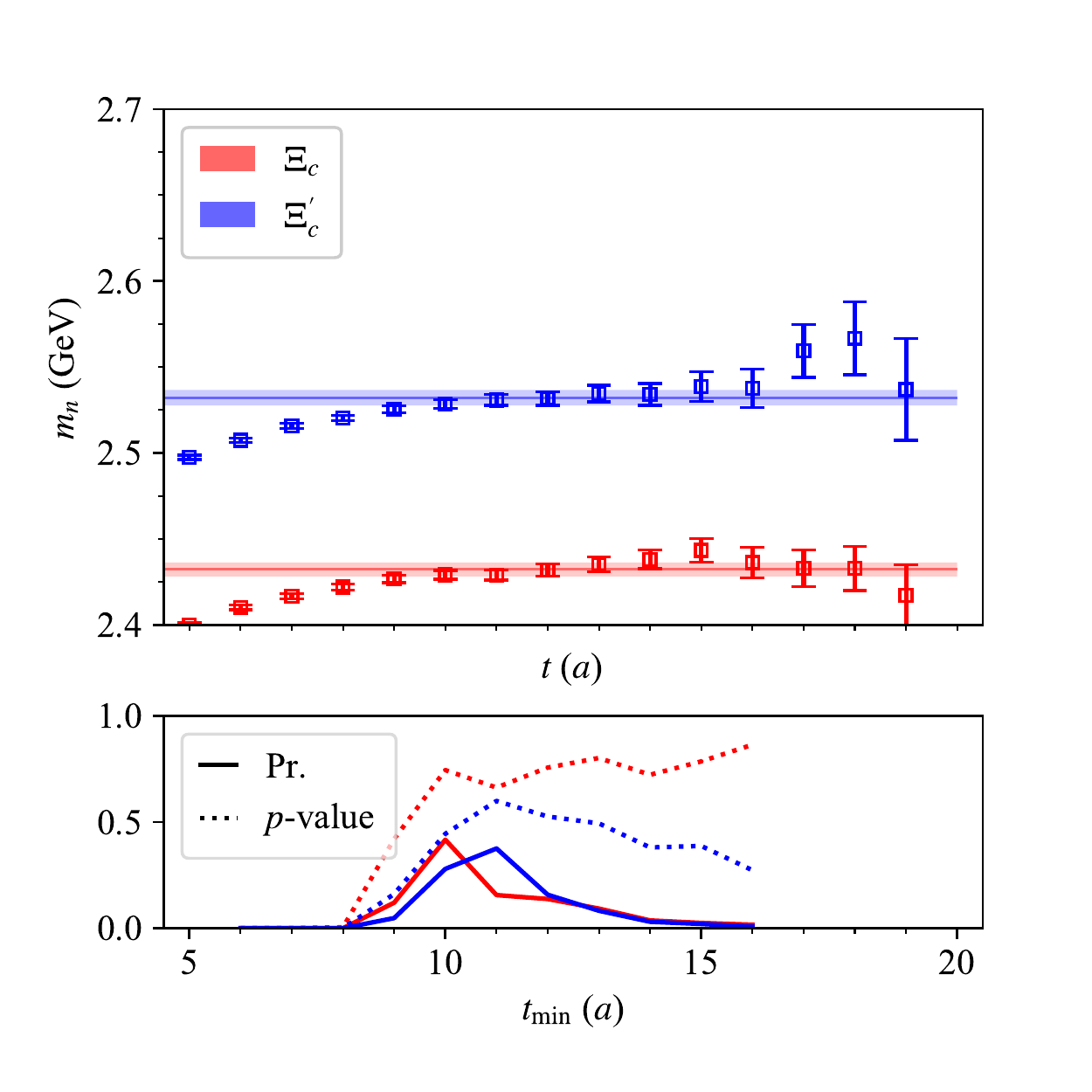}}
\subfigure[ ~	 C11P29S]{\includegraphics[width=0.45\textwidth]{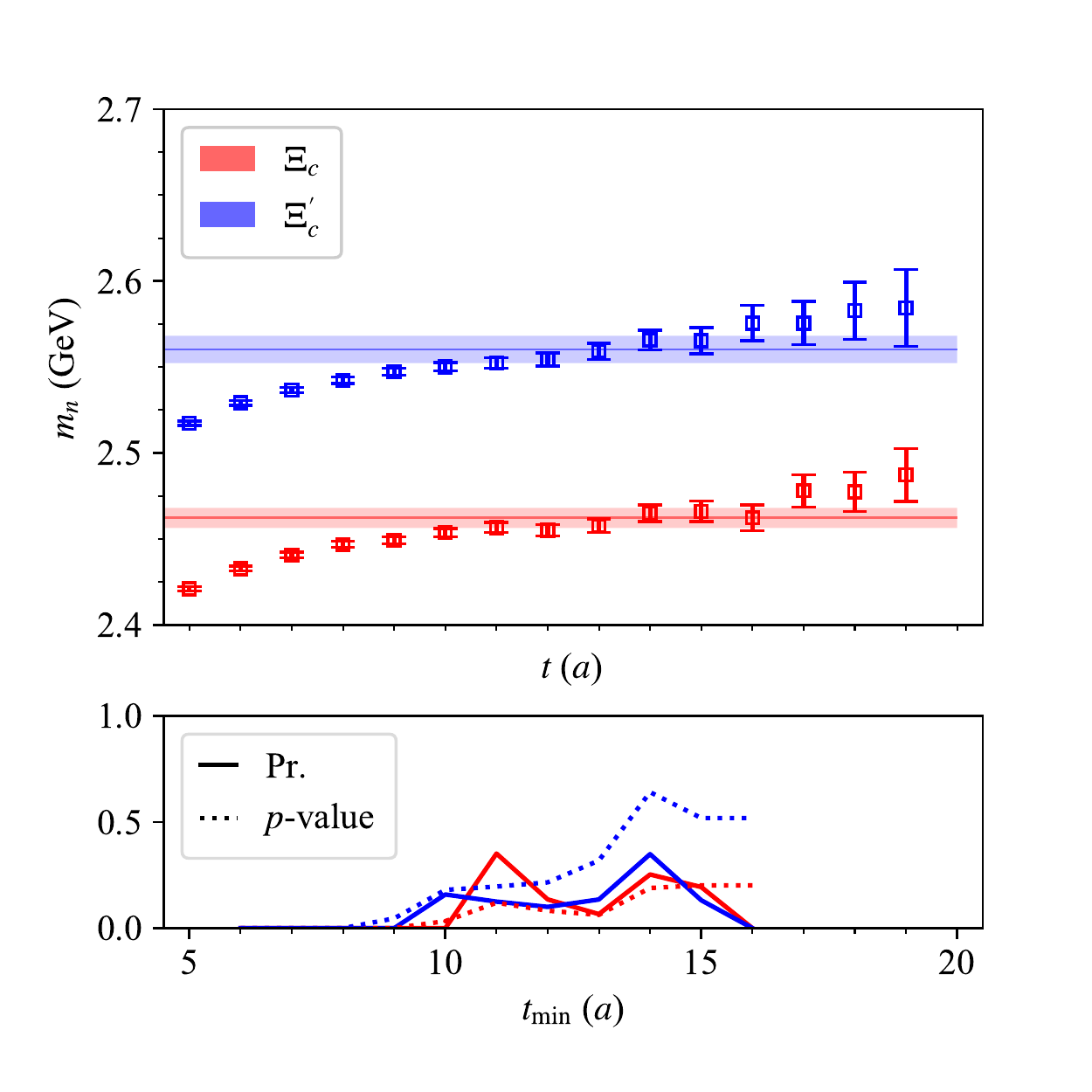}}
\subfigure[ ~	 C08P30S]{\includegraphics[width=0.45\textwidth]{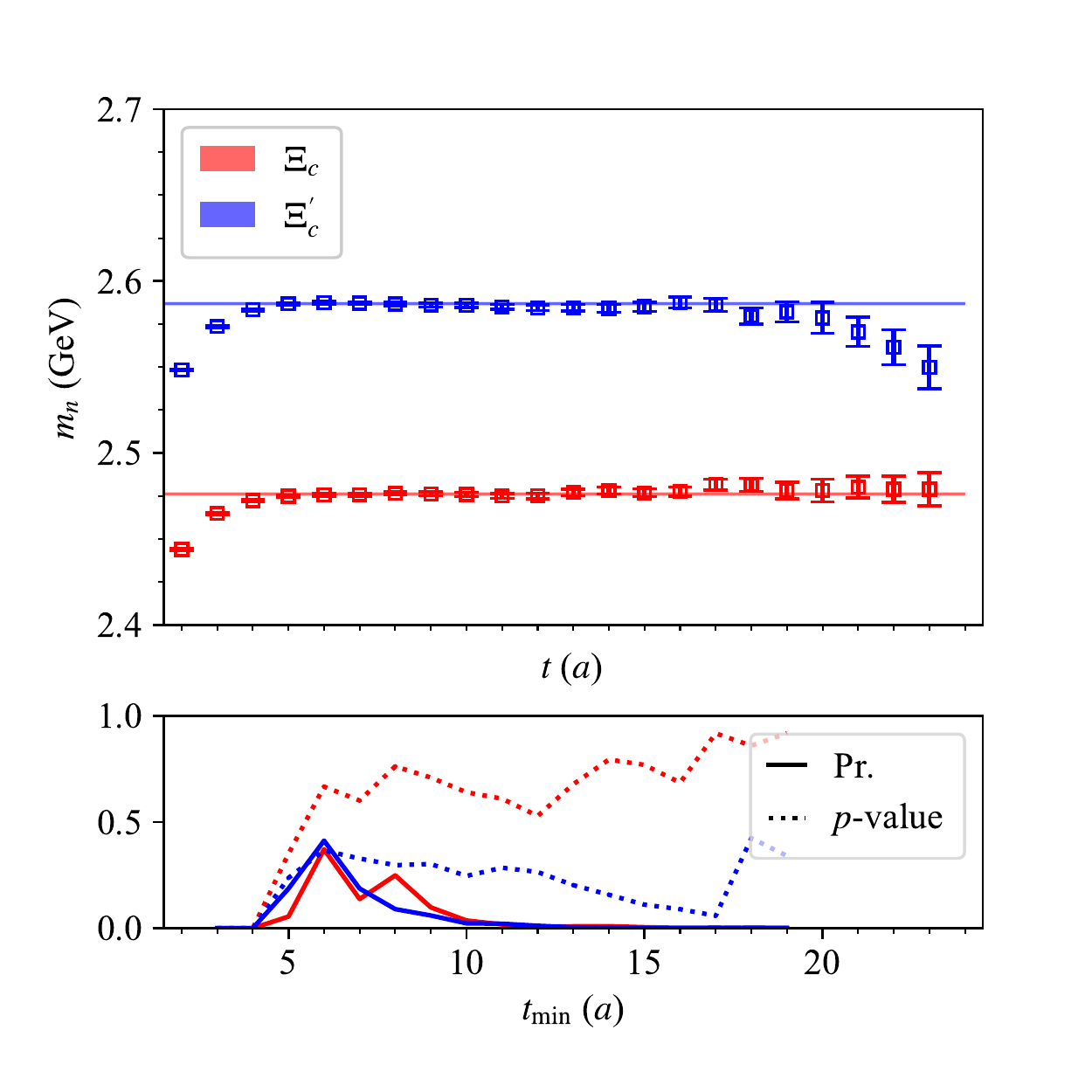}}
\subfigure[ ~	C06P30S]{\includegraphics[width=0.45\textwidth]{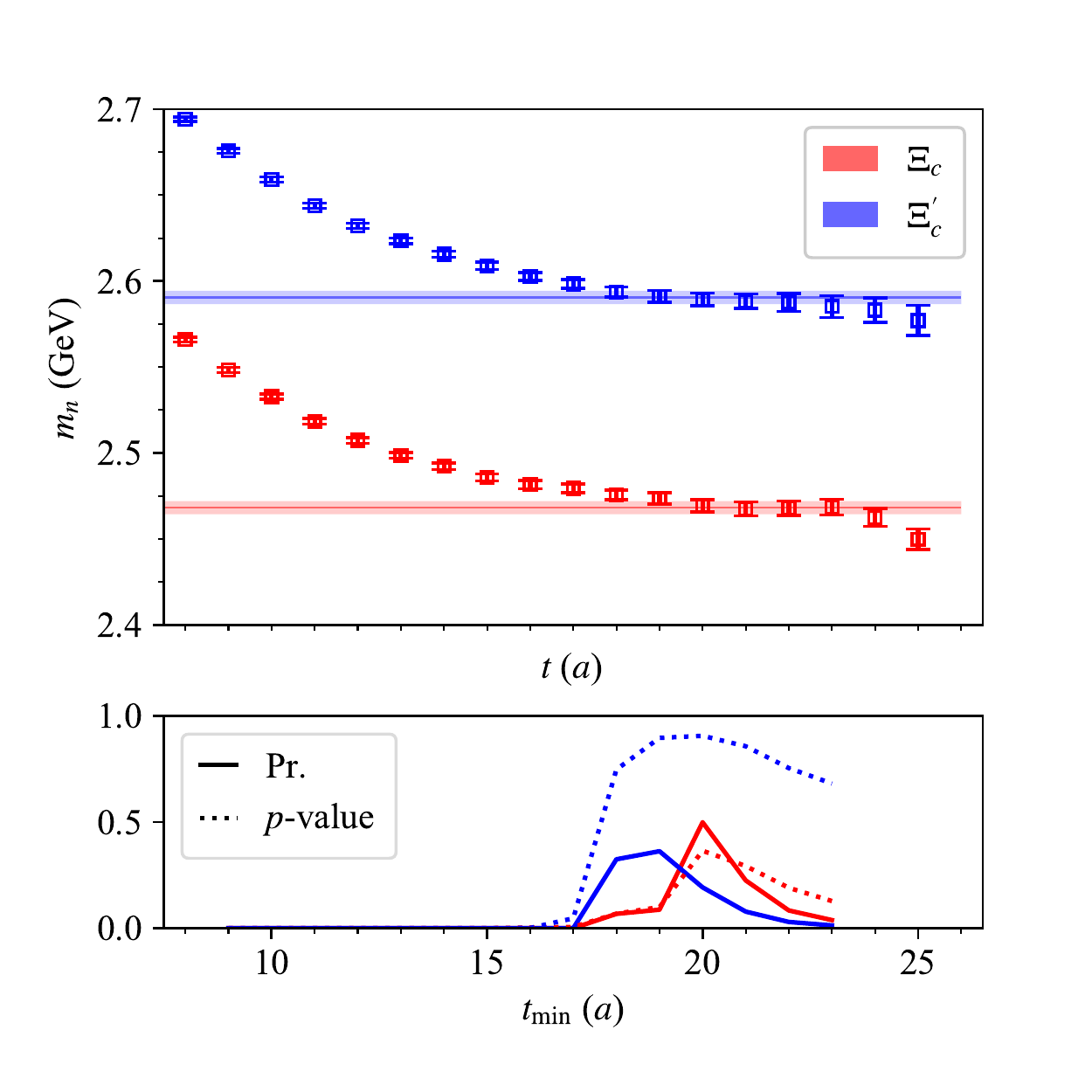}}
\caption{Upper panel: effective masses from eigenvalues $\lambda_n(t)$ ($n=\Xi_c,\Xi_c'$),  and model averaging results on each ensembles. Lower panel: the standard $p$-values (dashed lines) and weight factors (solid lines) which reflect the fit quality. }
\label{fig:spec_all_gevp}
\end{figure*}

In our analysis in Sec.~\ref{sec:GEVP}, we set $t_{\mathrm{min}}=[6, 17]a$ and $t_{\mathrm{max}}=20a$ for the C11 ensembles, $t_{\mathrm{min}}=[4, 20]a$ and $t_{\mathrm{max}}=24a$ for C08P30S, and $t_{\mathrm{min}}=[9, 24]a$ and $t_{\mathrm{max}}=28a$ for C06P30S. The illustration of fit results and fit quality annotated by normalized weight factor $\mathrm{Pr}$ and $p$-value are collected in Fig.\ref{fig:gevp_effmass} as well as Fig.\ref{fig:spec_all_gevp}.

\end{appendix}

\end{document}